\documentclass{article}
\usepackage[utf8]{inputenc}
\usepackage{amsmath}
\usepackage{xfrac}

\title{An Enlightening Derivation of so-called Fermi's Golden Rule in Quantum Mechanics with some new perspectives including Quasi Adiabatic Following}
\author{ M G Burt
\\
\\Department of Physics, Durham University, Durham, DH1 3LE, UK }

\date{23rd June 2020  }

\begin{document}

\maketitle

\begin{abstract}
    A novel and readily understandable derivation of the Golden Rule of time dependent perturbation theory is presented. The derivation is based on quasi adiabatic turning on of the perturbation reminiscent of that used, for instance, in some formal developments of scattering theory. The approximate energy conservation is expressed in terms of an intuitively and physically appealing Lorentzian line shape rather than the artificial, oscillatory $\sin(x)/x$ type line shape that appears in conventional derivations.  The conditions for the derivation's validity are compactly and conveniently expressed in the frequency/energy domain rather than in the usual time domain. The derivation also highlights how, along with approximate energy conservation, the transition rate approximately follows the variations in the square of the perturbation as one may expect in a quasi adiabatic regime. In the first instance, the quasi adiabatic turning on is achieved, as usual, by a single exponential time variation. But we demonstrate that the following of the square of the perturbation by the transition rate is more general and that one can derive the Golden Rule for a general slowly varying time dependent perturbation. This allows one to derive generalisations of the simple decay law, originally derived in the classic paper by Weisskopf and Wigner; a tutorial exposition of the essence of this classic work is provided. The Oppenheimer method for applying the Golden rule to problems, such as the electric field ionisation of atoms, in which the perturbing potential can also create the final states, is reviewed. No use of an energy gap condition is needed to derive our results on quasi adiabatic behaviour in contrast to the original derivations of the adiabatic theorem in quantum mechanics.
\end{abstract}

\section{Introduction}

In the study of quantum mechanics the formula for the transition rate in time dependent perturbation theory, often referred to as Fermi's Golden Rule, occupies a unique position in the development of the theory between the mysterious microscopic world  of quantum mechanics proper with its interference effects and the disturbance associated with observation and the more familiar semi-classical world of energy levels along with associated transition rates. It is also a difficult topic because it involves a `Goldilocks' type approach : the perturbation must not be applied for too short a time otherwise the constant transition rate is not established; on the other hand it must not be applied for too long otherwise the change in occupation probability of the initial state becomes too large for the perturbation assumption to hold. And yet, we often want to understand and describe decay processes in their entirety which includes instances for which the occupation probability of the initial state is not close to unity; every time one wants to apply the rule to a fresh time domain with a different initial state occupation probability one has to go through this `Goldilocks' balancing act, which is conceptually ugly to say the least.  Further we might want to consider problems in which the decay or transition rate is not constant, but varies with time such as tunnelling out of a potential well induced by a time dependent electric field.  The conventional presentation of time dependent perturbation theory in which a constant perturbation is abruptly turned on at time zero, does not show how to tackle this problem.

Indeed, the author has found the treatment of the Golden rule in textbooks generally less than ideal ( it would be invidious to give examples since the comment is a matter of taste rather than correctness ), even if one restricts oneself to the simple case of a constant perturbation leading to a constant transition rate. The derivation usually requires division by frequency differences which one knows must be small or even zero, thereby \textit{prima facie} undermining the usual justification for using perturbation theory. While it is possible to prove that the divisions by zero are not formally a problem$^1$, it is highly desirable to have a derivation in which this distraction is not present. And it is also desirable to have a derivation in which the constraints are not expressed in the time domain, so that within the limitations of the derivation, the transition rate derived is applicable to all times or at least a substantial continuous domain during which the initial state occupancy can vary appreciably. 

Another aspect of the conventional derivations of the Golden Rule is that, after sufficient time, the approximate constant transition rate and the approximate energy conservation are both achieved simultaneously. What is not highlighted is the fact that after the same time interval, if the perturbation is time dependent, the transition rate follows the square of the perturbation and that this following is  effectively instantaneous$^{2,3}$. So, when the Golden Rule is valid, the instantaneous transition rate does not depend on previous values of the perturbation. One of the purposes of this paper is to highlight this quasi adiabatic following as it highlights the pivotal position of the Golden Rule in the development of quantum theory.

Before introducing the contents of this paper, a comment on the use of the phrase `so called Fermi's Golden Rule' . It is easy to get the impression (and the author is included among those previously under this impression and was so for many years) that the Golden Rule formula was derived by Fermi himself. But that is not born out by the documentary evidence. Indeed, the author was alerted to this fact by reading the footnote to the term `Golden Rule' given in Sakurai's book$^4$ who referenced Fermi's lectures on nuclear physics$^5$. Fermi, while calling the transition rate formula `The Golden Rule', actually references Schiff's book$^1$ who in turn references Dirac$^{6,7}$ as the originator. ( This history has been summarised in a note by Visser $^8$.) To be fair on the authors who followed Dirac, the latter used energy normalisation for the basis functions of the final states, an elegant approach, but incidentally obscured the role the density of states plays in many applications. Nonetheless Dirac's formulation is universally applicable.

This paper starts  with a derivation of the Golden Rule that has great appeal as it keeps the constraints out of the time domain, avoids any possible division by zero, brings out the close connection between energy conservation and adiabaticity i.e. the quasi instantaneous following of the square of the perturbation by the transition rate. Our derivation has its origins in the formal theory of scattering ( see e.g. Merzbacher$^9$ )  in which one turns on the perturbation adiabatically, as a rising exponential, but our derivation does not need to take the strict adiabatic limit and so avoids the contradictions that that raises when trying to apply it to transition rates as opposed to just energy levels and wavefunctions. The close relation between the quasi adiabatic behaviour of the transition rate with respect to the time varying perturbation and the more conventional relation of the wavefunction to the same is pointed out. This derivation of the Golden Rule is expanded in the following section to both rising and falling exponential variations of the perturbation to provide evidence that this adiabatic following is not just limited to a single rising exponential confirming earlier work using a density matrix approach $^2$.

We then move on ( Section 4 )  to explore the approximate following for arbitrary, albeit sufficiently slowly varying, perturbations which leads on to our generalisation ( section 5 ) of the Weisskopf-Wigner simple exponential decay result$^{10}$, due to a constant perturbation, a generalisation in which the time constant varies with the perturbation.

The author has been perplexed and confused by the summaries of the famous Weisskopf-Wigner paper$^{10}$ he has seen, which have clearly been intended for a wide audience not interested in theoretical detail, so he has consulted the original paper which uses a very different approach to that used here. Since this classic paper uses some very subtle and valuable arguments, the essence has been distilled for the convenience of the reader in section 6. It will become clear that the method used by Weisskopf-Wigner does not have a ready generalisation to the case of a time dependent perturbation as treated by the author.

Up to this point, it has been assumed that the final states, at least in the first instance , are discrete albeit forming a quasi continuum. Nonetheless, it can be convenient, especially when the perturbation is an applied electric field, to follow Dirac$^7$ and assume, \textit{ab initio}, that the final states form a continuum. In section 7, we show how this can be done with energy normalisation of the final states. This leads on to our discussion of the Oppenheimer$^{11}$ formulation of time dependent perturbation theory as the time evolution of an initial non-stationary state; he developed it to tackle electric field ionisation of an atom in which the perturbation actually creates the final states as well as inducing transitions thereto. 

\section{A Simple but Enlightening Derivation of the Golden Rule }

Consider the transitions caused by introducing a time dependent perturbation, $V(t)$ , which rises exponentially in time. We will write $V(t)=V(0) \exp(\gamma t )=V \exp(\gamma t ) $ with $\gamma > 0$ so that the perturbation is zero in the distant past. In the absence of perturbation, the unperturbed system to which it is applied has orthonormal time independent stationary states $ \Psi_n $ with energies $ E_n $.The time dependent wavefunction, $ \Psi(t) $, is expanded with the time dependent coefficients, $c_n(t)$  as 
\begin {equation}
 \Psi(t) = \sum_n c_n(t) \exp{(-i E_n t / \hbar)}\Psi_n
\end  {equation}
The time dependent coefficients , $ c_n(t) $, obey the equation

\begin{equation}
 i \hbar \frac{dc_n(t) }{dt} = \sum_m V_{nm}(t)\exp{(-i (E_m - E_n) t / \hbar)}c_m(t) 
\end{equation}
or
\begin{equation}
 i \hbar \frac{dc_n(t) }{dt} = \sum_m V_{nm}(t)\exp{(+i \omega_{n,m} t )}c_m(t) 
\end{equation}
where
\begin{equation}
 \hbar \omega_{n,m}  = E_n - E_m 
\end{equation} and it has been convenient to have the matrix elements of the perturbation in the Schr{\"o}dinger picture so the only time variation is due to the perturbation itself.
 As usual we consider the unperturbed system in the distant past to be in some initial state, $i$,  so that $ c_i(-\infty ) = 1 $ and $ c_f(-\infty ) = 0  $ for all $ f $,  the final states,  those not equal to  $i$.
 
 In the usual way, we obtain the equations for the $c_f$ to the lowest order of accuracy by substituting the initial values of the $c$'s into the RHS of (3) to obtain
 \begin{equation}
 i \hbar \frac{dc_f(t) }{dt} \approx  V_{fi}(t)\exp{(+i \omega_{f,i} t )} 
\end{equation}
so that 

\begin{equation}
 c_f(t) \approx \frac{-i}{\hbar}  V_{fi} \frac{\exp{[(+i \omega_{f,i}+\gamma  ) t}]}{+i \omega_{f,i}+\gamma } 
\end{equation}
where there is manifestly no problem with possible divide by zero if $\omega_{f,i}=0$ because of the presence of real non-zero $\gamma$.
Straightforward manipulation gives

\begin{equation}
 \frac{d |c_f(t)|^2 }{dt} \approx \frac{2 \pi}{\hbar}  |V_{fi}|^2 \Delta ( E_{fi}, \Gamma ) e^{2\gamma t}
\end{equation}
where 
\begin{equation}
    \Delta (E,\Gamma) = \frac{1}{\pi} \frac{\Gamma}{E^2 + \Gamma^2}
\end{equation}
with $E_{fi}= \hbar \omega_{fi}$ and $\Gamma= \hbar \gamma$.
The reader will recognize that $\Delta$ is none other than a normalized Lorenztian line shape which is a possible representation of a $\delta$ function in the limit $\Gamma \rightarrow 0$. But we will not take this limit, as we do not need to, and, if one does, it just leads to complications as we will see in a moment.
Now just suppose, for simplicity , that $V_{fi}$ does not depend on the initial and final states, such as is the case for low energy scattering from a localised scatterer; let us denote $V_{fi}$ by $V_m$ to show that it is manifestly independent of the initial and final states yet differs from just the perturbation $V$ itself, an operator, $V_m$ being just a complex number.  Then from (7) the total transition rate, $r(t)$, out of the initial state is 

\begin{equation}
r(t) = \sum_f \frac{d |c_f(t)|^2 }{dt} \approx \frac{2 \pi}{\hbar}  |V_m|^2 \Bigg[\sum_f  \Delta ( E_{fi}, \Gamma )\Bigg]  e^{2\gamma t}.
\end{equation}

Now we suppose that the final states form a quasi continuum of levels approximately uniformly spaced over the line width about energy $E_i$ so we need

\begin{equation}
\Bigg| \frac{dD}{dE} \Bigg| \Gamma  \ll  D ( E ) 
\end{equation}
generally or at least for  $E \approx E_i$, where $D ( E ) $ is the number of states per unit energy often referred to as the density of states. We need the modulus sign in (10) since the derivative of the density of states can be negative, as for free particles moving in one dimension. If condition (10) is fulfilled, then it is straightforward to evaluate the sum in equation (9) because $D(E_i)$ is approximately constant across the Lorentzian line shape $\Delta ( E_{f}- E_{i}, \Gamma )$ : 
\begin{align}
 \sum_f  \Delta ( E_{fi}, \Gamma ) & = \int D(E_f) \Delta ( E_{f}- E_{i}, \Gamma ) \,\mathrm{d} E_f \nonumber \\
 & \approx D(E_i) \int  \Delta ( E_{f}- E_{i}, \Gamma ) \,\mathrm{d} E_f \nonumber \\
 & = D(E_i) 
\end{align}
the density of final states at the initial energy $E_i$ as one would expect from energy conservation. So, as long as the turn on has been sufficiently slow to ensure approximate energy conservation, we have, for the transition rate, $r$, from the initial state,

\begin{equation}
r(t)  \approx \frac{2 \pi}{\hbar}  |V_m|^2 D ( E_i )  e^{2\gamma t}
\end{equation}
or
\begin{equation}
r(t)  \approx \frac{2 \pi}{\hbar}  |V_m(t)|^2 D ( E_i ) . 
\end{equation}

We see that the transition rate at time $t$ follows the the modulus squared of the matrix element of the perturbation $V(t)$ at time $t$ quasi adiabatically and with a prefactor independent of $\gamma$. This suggests that, when a perturbation has been turned on slowly enough for approximate energy conservation to be achieved, all memory effects of the perturbation at earlier times are irrelevant. We are dealing with quasi adiabatic following. We will explore this more fully in a later section. But we can now say with some confidence that when the perturbation is $V$ and (10) holds, then the transition rate, $r$ is given by

\begin{equation}
r = r(0)  \approx \frac{2 \pi}{\hbar}  |V_m|^2 D ( E_i )  
\end{equation}
which is none other than the Golden Rule. From our derivation our result will hold for all times $t$ such that $\gamma t \ll 1 $ i.e while the perturbation has strength $V$.
If the perturbation approximation is to hold good, then we must have $c_i(t) \approx 1$. The extent to which $c_i(t) $ differs from unity is just
\begin{equation}
\Delta |c_i(t)|^2 = \int_{-\infty} ^{t} w(t') \mathrm{d} t'
\end{equation}
or from (13)
\begin{equation}
\Delta |c_i(t)|^2 \approx \frac{ \pi}{\Gamma}  |V_m(t)|^2 D ( E_i ).  
\end{equation}

We see from (16) the folly in taking the limit $\Gamma \rightarrow 0 $ literally. Eventually $\Delta |c_i(t)|^2 $ will approach, and even exceed, unity during this limit taking process, removing all justification for using perturbation theory, unless $V \rightarrow 0$ sufficiently fast as  $\Gamma \rightarrow 0 $ i.e. one goes to the limit of vanishingly small perturbation, a self defeating exercise.

We can now summarize the conditions for the validity of our derivation of the Golden Rule as
\begin {equation}
 \pi  |V_m(t)|^2 D ( E ) \ll \Gamma \ll \Bigg|\frac{dE}{d\ln[D(E)]} \Bigg|
 \end{equation}
 which is in a form that provides a ready test as to whether the use of the Golden Rule is appropriate. Of course, there are some cases in which $V$ and $\Gamma$ are within the control of the experimenter e.g the application of an electric field.
 
 Sometimes it is more convenient to have this condition in terms of rates and frequencies in which case we rewrite ( 17 ) as 
 
 \begin {equation}
 r/2  \ll \gamma \ll \Bigg|\frac{d\omega}{d\ln[D(\omega)]} \Bigg|.
 \end{equation}
 \\
 The density of states, of course, is not quite the same function of frequency as it is of energy, but for convenience $D(\omega)$ will be used for $\hbar D(E)$.
 
It is interesting to contrast the constraints (18) with those needed on the time for the validity of the conventional derivation, i.e. starting with the system in a definite initial state at time zero and then subjecting it to a constant perturbation, which are the constraints on the time, $t$, for a constant transition rate and yet minimal depletion of the initial state occupancy: 
 \begin {equation}
 r  \ll 1/t \ll \Bigg|\frac{d\omega}{d\ln[D(\omega)]} \Bigg|. \nonumber
 \end{equation}
 In contrast , the constraint (18) for our derivation is on $\gamma$, the rate of change of the perturbing potential, not the time. This suggests, along with the quasi adiabatic following and associated memory loss, that the theory developed here can be extended to arbitrary slowly varying perturbations without any restriction on the time and this is exactly what we will demonstrate in section 5.
\\
\\
When $V_{if}$ is not independent of $i$ and $f$, then the principal formulae derived in this section and conclusions deduced, are still valid, but $|V_m(t)|^2$ has to be replaced by a suitable average value of $|V_{if}(t)|^2$ , as shown in Appendix A. The reader will appreciate that $V_m$ may depend on the final state energy, but that the derivation can easily be modified to take this into account by treating the product  $|V_m(E_f,t)|^2 D(E_f)$ in the same way as $ D(E_f)$ ; we end up with $|V_m(E_f,t)|^2$ in the transition rate formula being evaluated at the initial state energy, $E_i$.
\\
\\
The extension of our derivation in this section to a harmonic time dependent perturbation is given in Appendix B.
\\
\\
Now, of course, the quasi adiabatic following highlighted here has only been shown for a single simple exponential variation.  However as shown in Appendix C the adiabatic following result (13) also holds when the time dependence of the perturbation takes the form of a superposition of exponentials which gives us added  confidence that it is generally applicable, with the associated memory loss, for slowly varying perturbations as we have just mentioned and will be confirmed in section 5.

Finally, we have been talking of the behaviour of the transition rate in following the square of the perturbation as adiabatic or more specifically quasi adiabatic. Normally when one speaks of the adiabatic approximation in quantum mechanics one is referring to constructing a stationary state wavefunction using the slow turn on of a potential. Are we talking about related things when we talk about transition rates? Indeed, we are. To see this we need to construct the wavefunction to first order using (6) to obtain

\begin{equation}
   \Psi(t) =   \Psi_i e^{-iE_it/\hbar}+  \sum_f\frac{ V_{fi} }{(E_i+i\Gamma)-E_f}\Psi_f  e^{-i(E_i+i\Gamma)t/\hbar} \nonumber.
\end{equation}
To obtain a simple stationary state time dependence $e^{-iE_it/\hbar}$ as required by energy conservation, we need $\Gamma \ll E_i$. This is essentially the same condition as in (10) and (17) since generally $D(E) \sim E^n$ and $n$ is of order unity.

The physics becomes clearer if one rewrites the equation for $\Psi(t)$ as 
\begin{align}
   \Psi(t) &=  \Bigg [ \Psi_i + \bigg ( \sum_f\frac{ V_{fi} }{(E_i+i\Gamma)-E_f}\Psi_f \bigg )e^{\Gamma t/\hbar} \Bigg ]  e^{-iE_it/\hbar} \nonumber  \\
   \text{or} \nonumber \\
   \Psi(t) &=  \Bigg [ \Psi_i +  \sum_f\frac{ V_{fi}(t) }{(E_i+i\Gamma)-E_f}\Psi_f  \Bigg ]  e^{-iE_it/\hbar} \nonumber .
\end{align}

Provided $\Gamma$ is sufficiently small, i.e. so that 
\begin{equation}
     \bigg ( \sum_f\frac{ V_{fi}(t) }{(E_i+i\Gamma)-E_f}\Psi_f \bigg ) \nonumber 
\end{equation}
is essentially independent of $\Gamma$ save for the $e^{\Gamma t/\hbar}$ factor in $V_{fi}(t)$, then  $\Psi(t)$ is a continuously evolving eigenstate of the total Hamiltonian at time $t$, an assertion at least accurate to first order in perturbation theory. On a much longer time scale the existence of the admixture of final states leads to transitions to those states at a rate determined by the value of the perturbing potential at that time.

Of course, the above first order perturbation theory expression for the wavefunction  with an exponentially varying perturbation is also valid no matter how the final state energy levels are distributed. If the final states are well separated in energy from the initial state, i.e. there is an energy gap, then the wavefunction will only depend on $\Gamma$ only via the perturbation itself (i.e. if $\Gamma \ll |E_f - E_i|$, usually a much milder restriction than (10), that for the continuous spectrum case) and we have adiabatic following, which corresponds to the traditional way of establishing the adiabatic theorem. But while an energy gap is a sufficient condition for adiabatic following, it is not necessary, at least for first order time dependent perturbation theory.

\section{Pulse with both Rising and Falling Exponential edges}
Our analysis of the transition rate due to a perturbation switched on as a rising exponential in the previous section has not only provided us with a simple derivation of the Golden Rule but also provided us with an insight into its potential validity for slowly varying, as opposed to just constant, perturbations generally. The fact that the transition rate followed the time dependence for the square of the perturbation with a constant of proportionality independent of the rate constant in the exponential suggests that the result is more general. To test this idea and explore it more generally, we now consider the transition rate for a pulse with an exponential for both the leading and trailing edge.This has been done earlier by the author $^2$ using a more complex treatment based on the density matrix with a perturbation varying sinusoidally in time and also with the added complication of phenomenological relaxation constants. The treatment given here is much more straightforward. We consider the perturbation potential 

\begin{align}
 V(t) &= V^{(-)}\exp{(+\gamma^{(-)}t)}    && t<0 \nonumber \\
  V(t) &= V^{(+)}\exp{(-\gamma^{(+)}t)}    && t>0    
\end{align}
where both $\gamma^{(\pm)}$ obey the constraint (18) and we do not need to be specific as to the value of $V(t)$ at $t=0$; indeed, there is no problem with having $V^{(-)} \ne V^{(+)}$ and we will assume this is the case . For the transition rate for $t<0$ we only need to make notational changes in equations (12) and (13) from section 2 to obtain
\begin{equation}
r(t)  \approx \frac{2 \pi}{\hbar}  |V^{(-)}_m|^2 D ( E_i )  e^{2\gamma t}
\end{equation}
or
\begin{equation}
r(t)  \approx \frac{2 \pi}{\hbar}  |V^{(-)}_m(t)|^2 D ( E_i ) . 
\end{equation}
The equation of motion for the final state amplitude, $c_f$, for $t>0$ is, assuming the pulse to be sufficiently weak that $c_i\approx 1$ throughout
\begin{equation}
 i \hbar \frac{dc_f(t) }{dt} \approx  V^{(+)}_{fi}(t)\exp{(+i \omega_{f,i} t )} 
\end{equation}
or
\begin{equation}
 i \hbar \frac{dc_f(t) }{dt} \approx  V^{(+)}_{fi}\exp{[+i( \omega_{f,i} + i\gamma^{(+)}) t ]}. 
\end{equation}
Integrating from $0$ to $t$ we obtain
\begin{equation}
 c_f(t) \approx  c_f(0) + \frac{-1}{ \hbar} V^{(+)}_{fi}\frac{\exp{[+i( \omega_{f,i} + i\gamma^{(+)}) t ]}- 1}{\omega_{f,i} + i\gamma^{(+)}} 
\end{equation}
where $c_f(0)$ is given by equation (6) with updated notation i.e.
\begin{equation}
 c_f(0) \approx \frac{-1}{\hbar}  V^{(-)}_{fi} \frac{1}{ \omega_{f,i}-i\gamma^{(-)} } 
\end{equation}
Now only the $c_f(0)$ term on the RHS of (24) contains information about the rising edge of the pulse ( via $V^{(-)}$ and $\gamma^{(-)} $) and we need to investigate its influence on the overall transition rate, $\sum_f \sfrac{d|c_f|^2}{dt}$, to the final states. To this end we will use
\begin{equation}
    \frac{d|c_f|^2}{dt} = c_f(\frac{dc_f}{dt})^* + c.c.
\end{equation}
It is clear from equation (23) that $\sfrac{dc_f}{dt}$ for positive times will only contain information on the falling part of the pulse so that the only contribution to the overall transition rate for the falling edge due to the rising edge will be contained in 
\begin{equation}
    \sum_f c_f(0) (\frac{dc_f}{dt})^* = \frac{-i}{\hbar^2}\exp{(-\gamma^{(+)}t)}\sum_f   V^{(-)}_{fi}V^{(+)*}_{fi} \frac{1}{ \omega_{f,i}-i\gamma^{(-)} } \exp{(-i \omega_{f,i} t )}
\end{equation}
and its complex conjugate. Making the nonessential, but notationally convenient, assumption that the matrix elements of the potential do not depend on $f$, the sum in the RHS of (27) becomes
\begin{equation}
\sum_f \frac{1}{ \omega_{f,i}-i\gamma^{(-)} } \exp{(-i\omega_{f,i} t )}
\end{equation}
which we can express,using the density of states, as an integral, $I$,  with respect to the variable $x = \sfrac{\omega_{f,i}}{\gamma^{(-)}}$
\begin{equation}
I = \int D(\omega_i + x \gamma^{(-)} ) \frac{1}{ x-i } \exp{(-ix \gamma^{(-)} t )}\mathrm{d}x
\end{equation}
If $\gamma^{(-)}$ is sufficiently small, then $x \gamma^{(-)} $ will be negligible compared with $\omega_i$ except for large $x$ for which the $\sfrac{1}{ (x-i) } $ factor will be small and, combined with the oscillating exponential, will give negligible contribution to the integral . In which case there is little error in evaluating $D$ at $\omega_i$ and taking it outside the integral. So,
\begin{equation}
I =  D(\omega_i)\int_{-\infty} ^{+\infty} \frac{1}{ x-i } \exp{(-ix \gamma^{(-)} t )}\mathrm{d}x
\end{equation}
where, in setting the limits, we have assumed that the initial state frequency, $\omega_i$, is not near the edge of the range of $D(\omega)$. This integral is easily evaluated, by closing the contour in the lower half plane ( since $\gamma^{(-)} t > 0$ for $t > 0 $) and using Jordan's lemma, to give zero since the integrand is analytic in the lower half plane, the only pole being at $x= + i $ in the upper half plane. This conclusion also applies to the complex conjugate of (27). So for sufficiently small $\gamma^{(-)}$ there is no influence on the overall transition rate during the falling edge due to the parameters determining the rising edge of the pulse. One notes that this is quite a robust approximation since, as $\gamma^{(-)}$ increases, thereby restricting the range of $x$ for which $D$ may be viewed as constant, the rate of oscillation of the exponential , determined by $\gamma^{(-)}t$, and hence the cancellation it provides, increases.

The reader will note that we have assumed $\gamma^{(-)} \ll \omega_i $ rather than 
\begin {equation}
  \gamma \ll \Bigg|\frac{d\omega}{dln[D(\omega)]} \Bigg|
 \end{equation}
 from (18). But typically $D(\omega) \sim \omega^n$ where $n$ is of order unity, so the conditions are not materially different.
 
Having shown that the $c_f(0)$ term on the RHS of (24)  makes no significant contribution, for small $\gamma^{(-)} $, to  the overall transition rate, $\sum_f \sfrac{d|c_f|^2}{dt}$, to the final states we need to evaluate the latter. Using (24) and (26) we have

\begin{multline}
\sum_f\frac{d|c_f|^2}{dt} \approx  \frac{-i}{ \hbar^2} |V^{(+)}_m|^2 \sum_f \Bigg[\frac{-2i\gamma^{(+)}}{\omega_{f,i}^2 + \gamma^{(+)2}}\exp{[-2\gamma^{(+)} t ]} \\ 
+ \bigg(\frac{\exp{[+i \omega_{f,i}t ]}}{\omega_{f,i} - i\gamma^{(+)}} - \frac{\exp{[-i \omega_{f,i}t ]}}{\omega_{f,i} + i\gamma^{(+)}}\bigg)\exp{[ -\gamma^{(+)} t ]}\Bigg]  
\end{multline}
 where again we have made the nonessential, but notationally convenient, assumption that the matrix elements of the potential do not depend on $f$. In converting this to an integral with the help of the density of states, $D(\omega_f)$, we note that for sufficiently small $ \gamma^{(+)} $  the integrand is sharply peaked around $ \omega_{f,i} \approx 0$. If the density of states is sufficiently slowly varying , i.e. (18) holds , then we can take $D(\omega_f)$ outside the integral and evaluate it at $\omega_i$ and extend the range of integration to all  $\omega_f$ to obtain
 \begin{multline}
\sum_f\frac{d|c_f|^2}{dt} \approx  \frac{-i}{ \hbar^2} |V^{(+)}_m|^2 D( \omega_i)\int_{-\infty} ^{+\infty} \Bigg[\frac{-2i\gamma^{(+)}}{\omega_{f,i}^2 + \gamma^{(+)2}}\exp{[-2\gamma^{(+)} t ]} \\ 
+ \bigg(\frac{\exp{[+i \omega_{f,i}t ]}}{\omega_{f,i} - i\gamma^{(+)}} - \frac{\exp{[-i \omega_{f,i}t ]}}{\omega_{f,i} + i\gamma^{(+)}}\bigg)\exp{[ -\gamma^{(+)} t ]}\Bigg]  \mathrm{d}\omega_f
\end{multline}
Carrying out the integration one obtains
\begin{equation}
r(t)  \approx \frac{2 \pi}{\hbar}  |V^{(+)}_m|^2 D ( E_i )  e^{-2\gamma^{(+)} t}
\end{equation}
or
\begin{equation}
r(t)  \approx \frac{2 \pi}{\hbar}  |V^{(+)}_m(t)|^2 D ( E_i )  
\end{equation}
and we see that the transition rate out of the initial state indeed follows the square of the perturbation during the trailing edge just as it did for the rising edge if the turn off and turn on of the pulse are each sufficiently slow. In addition the perturbation must be sufficiently weak so as not to deplete the initial state significantly i.e. so that the  condition (18) is fulfilled for each part, both rising and falling, of the pulse separately.

In appendix C we have shown that the conclusions of section 2, that the transition rate depends on time as the square of the perturbation, hold not only for a single rising exponential, but for any linear combination of such. The reader will be able to carry out a similar analysis to show that the same and extended conclusions of this section are applicable to independent linear combinations of exponentials for both the the rising and falling edges.

So we have now confirmed our suspicions that our Golden Rule derivation for slowly varying perturbations is not restricted to just a rising exponential or combinations there of, but is valid for a complete pulse, albeit a rather artificial one. We now move on in the next section to generalise our result to a train of pulses of an arbitrary shape.

\section{`Pulse' Independence}
In the previous section we have seen that the transition rate in the trailing edge of the pulse is approximately independent of the parameter $\gamma^{(-)}$  defining the leading edge provided 
\begin {equation}
  \gamma^{(-)} \ll \Bigg|\frac{d\omega}{dln[D(\omega)]} \Bigg|
 \end{equation}
 i.e. one is in the quasi adiabatic regime. We note further that there is no constraint on the relative sizes of $ V^{(\pm)}$ , so the results hold even if there were a discontinuity at $t = 0$. If one considers the pulse in the previous section to be made up of two separate pulses, the first a rising exponential that drops to zero at $t=0$ and then the second rising abruptly at $t=0$ and then falling exponentially, then our result is saying that the effects in terms of the induced transition rates of the two pulses are independent. While it is perfectly possible to develop the method in the previous section to investigate generalisations of the Golden Rule, the author has found it simpler to focus on the overall transition probability (rather than the transition rate) out to the initial state into the final states for a pulse, establish the independence between pulses and then model any quasi adiabatic variation of a perturbation as a series of pulses.
 
 We shall also assume that the matrix elements of the perturbation between final states can be ignored so  $V_{ff'} =0$ , an appropriate assumption for many decay of excitation problems and weak scattering.
 
\subsection{Transition Probabilities for Successive Pulses are Additive}

Suppose we have a time dependent perturbation $V(t)$ in the form of a pulse. The equation of motion for the final state amplitude $c_f$ is
 \begin{equation}
 i \hbar \frac{dc_f(t) }{dt} =  V_{fi}(t)\exp{(+i \omega_{f,i} t )} c_i (t)
\end{equation}
We assume that the pulse is weak and does not alter $c_i$ much so that $c_i$ is approximately constant over the duration of the pulse and may as well be evaluated at the centre of the pulse for definiteness. So the change $\Delta c_f$ in $c_f$ due to the pulse is given by

\begin{equation}
 \Delta c_f \approx \frac{-i}{\hbar}\Bigg[ \int_{-\infty} ^{+\infty}V_{fi}(t)\exp{(+i \omega_{f,i} t )}\mathrm{d}t\Bigg]  c_i
\end{equation}
 The quantity $c_f$ only changes during the pulse, but we can extend the integration to $\pm \infty$ because this does not change the value of the integral. Introducing the Fourier transform $\tilde V(\omega)$ of $V(t)$ via
\begin{equation}
V(t)= \int_{-\infty} ^{+\infty}\frac{d\omega}{2 \pi}\tilde V(\omega)\exp{(-i\omega t)}
\end{equation}
to find 
\begin{align}
  \Delta c_f &\approx \frac{-i}{\hbar}\Bigg[\int_{-\infty} ^{+\infty}\frac{d\omega}{2 \pi}\tilde V_{fi}(\omega) \int_{-\infty} ^{+\infty}\exp{(+i( \omega_{f,i}-\omega )t )}\mathrm{d}t\Bigg]  c_i   \nonumber \\
  &= \frac{-i}{\hbar}\Bigg[\int_{-\infty} ^{+\infty}d\omega\tilde V_{fi}(\omega)  \delta( \omega_{f,i}-\omega )\Bigg]  c_i \nonumber \\
  &= \frac{-i}{\hbar}\tilde V_{fi}(\omega_{fi})  c_i .
\end{align}

Before going further it is worth pointing out that , within our model ($V_{ff'} =0$) equation (37) is exact and hence the result (40) depends only on our limitation that $c_i $ is little changed by the pulse.

Now, if we have two consecutive but separate non-overlapping pulses $V^{(1)}(t)$ and $V^{(2)}(t-T)$ with centres separated by time $T$, it follows that the total change in $c_f$ due to both pulses is given by 
\begin{equation}
\Delta c^{(1 and 2)}_f =\frac{-i}{\hbar}\Bigg[\tilde V^{(1)}_{fi}(\omega_{fi} ) c^{(1)}_i +\tilde V^{(2)}_{fi}(\omega_{fi} )c^{(2)}_i \exp{(i\omega_{fi}T)}  \Bigg] 
\end{equation}
where $c^{(1)}_i$ and $c^{(2)}_i$ are the approximate constant values of $c_i$ during the first and second pulses respectively. So the squared modulus of this change in $c_f$ is
\begin{align}
|\Delta c^{(1 and 2)}_f|^2 &=\frac{1}{\hbar^2}\Bigg|\tilde V^{(1)}_{fi}(\omega_{fi} ) c^{(1)}_i +\tilde V^{(2)}_{fi}(\omega_{fi} )c^{(2)}_i \exp{(i\omega_{fi}T)}  \Bigg|^2 \nonumber \\
&= |\Delta c^{(1)}_f|^2 +|\Delta c^{(2)}_f|^2 + (\Delta c^{(1)*}_f\Delta c^{(2)}_f \exp{(i\omega_{fi}T)} + c.c. )
\end{align}
Now the term $\Delta c^{(1)*}_f\Delta c^{(2)}_f \exp{(i\omega_{fi}T)}$ and its complex conjugate will be nonzero in general. But the exponential term within it, $\exp{(i\omega_{fi}T)}$, will be a rapidly varying function of $\omega_{fi}$ for large $T$. So provided the $\omega_{fi}$ dependent terms in $\Delta c_f^{(1)}$ and  $\Delta c_f^{(2)}$, namely $\tilde V^{(1)}_{fi}(\omega_{fi} )$ and $\tilde V^{(2)}_{fi}(\omega_{fi} )$, are relatively slowly varying functions of $\omega_{fi}$ we can expect the term $\Delta c^{(1)*}_f\Delta c^{(2)}_f \exp{(i\omega_{fi}T)}$ and its complex conjugate to give approximately zero on summation over final states.[The reader may object that there will be cases for which  $\omega_f \approx \omega_i $ so that this argument is flawed. It is just this problem that led the original proofs of the adiabatic theorem $^{11,12,13}$ , to which our interpulse independence i.e. adiabaticity is closely linked, to insist on an energy gap rather than unbroken continuum. But our results from previous sections and subsection 4.2.3 suggest that this in unnecessary, at least for transition rates within perturbation theory.] Supposing that $\sum_f \Delta c^{(1)*}_f\Delta c^{(2)}_f \exp{(i\omega_{fi}T)}$ (and consequently its complex conjugate) to be approximately zero, ( some examples will be given in the next subsection ) we may write
\begin{equation}
\sum_f|\Delta c^{(1 and 2)}_f|^2 \approx \sum_f|\Delta c^{(1)}_f|^2 +\sum_f|\Delta c^{(2)}_f|^2 .
\end{equation}
Now suppose further that in the distant past the system was in an initial state. Then $c_f =0$ initially and $\Delta c^{(1and2)}_f$ will just be $c^{(1and2)}_f$ the value of $c_f$ after both pulses 1 and 2. So, 
\begin{equation}
\sum_f|c^{(1 and 2)}_f|^2 =\sum_f|\Delta c^{(1)}_f|^2 +\sum_f|\Delta c^{(2)}_f|^2 .
\end{equation}
We can easily extend this result to any number of non-overlapping pulses by extending the argument for two pulses to an arbitrary number of pulses. Or more easily by just applying the two pulse result to three pulses by regarding pulses 1 and 2 together as one pulse. So from (26) we have
\begin{align}
\sum_f|c^{(1 and 2 and 3)}_f|^2 &= \sum_f|c^{((1 and 2) and 3)}_f|^2  \nonumber \\
&=\sum_f|\Delta c^{(1and2)}_f|^2 +\sum_f|\Delta c^{(3)}_f|^2
\nonumber \\
&=\sum_f|c^{(1and2)}_f|^2 +\sum_f|\Delta c^{(3)}_f|^2
\nonumber \\
&=\sum_f|\Delta c^{(1)}_f|^2 +\sum_f|\Delta c^{(2)}_f|^2 +\sum_f|\Delta c^{(3)}_f|^2 .
\end{align}
We can obviously carry out this procedure for as many pulses as we like or use the fact that for $n+1$ pulses we could take the first $n$ pulses as the first pulse and the $(n+1)\text{th}$ pulse as the second pulse and proceed to establish the general result by induction.
For our derivation of the generalised exponential decay law in the next section it is best to express our result as the change, $\Delta^{(n)}\big[\sum_f| c_f|^2\big]$, in $\sum_f |c_f|^2$ as a result or the $n\text{th}$ pulse, i.e.
\begin{equation}
\Delta^{(n)}\bigg[\sum_f| c_f|^2 \bigg] =\sum_f|\Delta c_f^{(n)}|^2
\end{equation}
where $\Delta c_f^{(n)}$ is the change in the amplitude $c_f$ due to the $n\text{th}$ pulse in isolation except for the factor of the reduced occupation probability of the initial state brought about by previous pulses. The only memory effect or legacy of previous pulses is the reduced occupation probability of the initial state.

\subsection{Examples :}
The term we have neglected in our derivation of (43) is
the "cross" or inter-pulse term
\begin{equation}
\sum_f \Delta c^{(1)*}_f\Delta  c^{(2)}_f \exp{(i\omega_{fi}T)}=\frac{c^{(1)*}_i c^{(2)*}_i}{\hbar^2}\sum_f \tilde V^{(1)}_{fi}(\omega_{fi} )^* \tilde V^{(2)}_{fi}(\omega_{fi} )\exp{(i\omega_{fi}T)} 
\end{equation}
and we need to investigate the circumstances under which the exponential term gives effective cancellation. To this end introduce the common shape $s(t)$ of each pulse with its Fourier transform $\tilde s(\omega)$ so that 
\begin{equation}
    V^{(n)}_{fi}(t) = V^{(n)}_{fi}(0)s(t)= V^{(n)}_{fi}s(t) 
\end{equation}
and write
\begin{align}
\sum_f \Delta c^{(1)*}_f\Delta  c^{(2)}_f \exp{(i\omega_{fi}T)}&=\frac{c^{(1)*}_i c^{(2)*}_i}{\hbar^2}\sum_f V^{(1)*}_{fi}V^{(2)}_{fi}|\tilde s(\omega_{fi} )|^2
\exp{(i\omega_{fi}T)}  \nonumber \\ 
&=c^{(1)*}_i c^{(2)*}_i\frac{V^{(1)*}_m V^{(2)}_m}{\hbar^2}\sum_f |\tilde s(\omega_{fi} )|^2 \exp{(i\omega_{fi}T)} \nonumber \\ 
&=c^{(1)*}_i c^{(2)*}_i\frac{V^{(1)*}_m V^{(2)}_m}{\hbar^2}I
\end{align}
where
\begin{equation}
    I = \int D(\omega_f)
|\tilde s(\omega_{fi} )|^2 \exp{(i\omega_{fi}T)}d\omega_f .
\end{equation}
Again we have made the nonessential, but notationally convenient, assumption that the dependence of the matrix element $V_{if}$ on $f$ can be ignored.
Note that (49) evaluated at  $T= 0$ gives the value of $I$ for the intra-pulse terms that are retained.

\subsubsection{Pulse with both Rising and Falling Exponential edges}
For our first example we return to the model used to extend our demonstration of adiabatic following in section 3 , but make the simplifying assumption that $V^{(-)} = V^{(+)}$. We find that
\begin{equation}
    |\tilde s(\omega)|^2=\frac{(\gamma^{(+)}+\gamma^{(-)})^2}{(\omega^2+\gamma^{(-)2})(\omega^2+\gamma^{(+)2})} .
\end{equation}
From (49) and (50) we find
\begin{equation}
    I = \int D(\omega_f)
\frac{(\gamma^{(+)}+\gamma^{(-)})^2}{(\omega_{fi}^2+\gamma^{(-)2})(\omega_{fi}^2+\gamma^{(+)2})} \exp{(i\omega_{fi}T)}d\omega_f
\end{equation}
For small $\gamma^{(\pm)}$ the integrand is strongly peaked at $\omega_{fi} = 0 $ and falls off as $\omega_{f,i}^{-4}$ at large 
$\omega_{f,i}$. We can take the $D(\omega_f)$ factor outside the integral and evaluate at $\omega_i$ and extend the limits of integration to $\pm\infty$ with little error.  So, 
\begin{equation}
    I \approx D(\omega_i)\int ^{+\infty}_{-\infty} 
\frac{(\gamma^{(+)}+\gamma^{(-)})^2}{(\omega_{fi}^2+\gamma^{(-)2})(\omega_{fi}^2+\gamma^{(+)2})} \exp{(i\omega_{fi}T)}d\omega_f
\end{equation}
Closing the contour in the upper half plane gives 
\begin{equation}
    I =  \frac{\pi}{\gamma^{(+)}-\gamma^{(-)}} \Bigg[\frac{e^{-\gamma^{(-)}T}}{\gamma^{(-)}}-\frac{e^{-\gamma^{(+)}T}}{\gamma^{(+)}}\Bigg]
\end{equation}
The value of $I$ for the intra-pulse terms, those retained, corresponds to $T=0$ or 
\begin{equation}
    I = \frac{\pi}{\gamma^{(+)}\gamma^{(-)}}
\end{equation}
showing that the discarded inter-pulse terms are of order $\exp{(-\gamma^{(\pm)}T)}$ smaller than those retained and this will be small for well separated pulses.

\subsubsection{Gaussian pulse}
In this case we have 
\begin{equation}
    V(t) = V (\tau \sqrt{\pi})^{-1} \exp{[-(t/\tau)^2]}
\end{equation}
so that
\begin{equation}
    s(t) = (\tau \sqrt{\pi})^{-1} \exp{[-(t/\tau)^2]}
\end{equation}
and
\begin{equation}
    |\tilde s(\omega )|^2=  \exp{[-(\frac{\omega \tau}{\sqrt{2}} )^2]}
\end{equation}
and
\begin{equation}
    I = \int D(\omega_f)   \exp{[-(\frac{\omega_{fi} \tau}{\sqrt{2}} )^2]} \exp{(i\omega_{fi}T)}d\omega_f .
\end{equation}
The presence of the gaussian in the integrand means that most of the contribution to the integral comes from $\omega_f$ values within a few $1/\tau$ of $\omega_i$. Provided $D(\omega_f)$ is approximately constant over this range we can evaluate it at $\omega_i$ and take it outside the integral and extend the limits to $\pm \infty$ so that

\begin{align}
    I &\approx  D(\omega_i)  \int ^{+\infty}_{-\infty}  \exp{[-(\frac{\omega_{fi} \tau}{\sqrt{2}} )^2]}
    \exp{(i\omega_{fi}T)}d\omega_f \nonumber \\
     I &=D(\omega_i) \frac{\sqrt{2 \pi}}{\tau} \exp{[-\frac{1}{2}(\sfrac{T}{\tau} )^2]}.
\end{align}
The corresponding integral for the retained terms can be evaluated by putting $T=0$ in the above expression showing that the discarded terms are, in order of magnitude,
\begin{equation}
\exp{[-\frac{1}{2}(\sfrac{T}{\tau})^2]}
\end{equation}
times those retained. Given that we must have (38) small for non-overlapping pulses we see that the discarded terms are
indeed negligible compared with those retained.

\subsubsection{Rectangular pulse}
This case essentially treats the usual constant perturbation case and generalises to the piece-wise constant perturbation case. Here we have
\begin{align}
    V(t) &= V   \nonumber &&\text{$ |t| < T/2$ } \\
         &= 0    &&\text{$ |t|> T/2 $ } 
\end{align}
so that
\begin{equation}
    s(t) = \theta^{(+)}(t+T/2) - \theta^{(+)}(t-T/2)
\end{equation}
and
\begin{align}
    |\tilde s(\omega_{fi} )|^2  &= \Bigg| \frac{\exp{(+i\omega_{fi}T/2)} - \exp{(-i\omega_{fi}T/2)}}{i \omega_{fi}} \Bigg|^2 \nonumber \\ 
    &= 4 \Bigg|\frac {\sin{[\omega_{fi}T/2]}}{\omega_{fi}}\Bigg|^2  \nonumber \\ 
    &\approx 2\pi T \delta(\omega_{fi}) &&  \text{for large T.}
\end{align}

As a digression, here is an example of a possible unsettling divide by zero, as one meets in the conventional derivation of the Golden Rule. It is not a problem, however, since taking the limit $\omega_{f} \rightarrow \omega_{i}$ in the first two lines of (64) gives the same answer as evaluating (64) afresh at $\omega_{fi} = 0$ as rigour requires: $\tilde s(0) = \int_{-T/2}^{+T/2}dt = T $ while $\lim_{\omega_{f} \rightarrow \omega_{i}}\tilde s(\omega_{fi} )$ gives the same answer, namely, $T$. This stability to potential divide by zero is a general feature of the time dependent perturbation series$^1$ , but the general case is hardly an elementary point. 
Continuing, we now have
\begin{equation}
    I = \int D(\omega_f) \Bigg| \frac{\exp{(+i\omega_{fi}T/2)} - \exp{(-i\omega_{fi}T/2)}}{i \omega_{fi}} \Bigg|^2  exp{(i\omega_{fi}T)}d\omega_f
\end{equation}
We now change variable to   $x= \omega_{fi} T$ so that

\begin{equation}
    I =-T \int D(\omega_i + x/T ) \frac{[ \exp{[2ix]}-2\exp{[ix]} +  1]}{x^2} dx
\end{equation}
Now $T$ is chosen so that only for large $x$ will  $D(\omega_i + x/T )$ differ appreciably from $D(\omega_i )$. The  $x^2$ in the denominator will ensure that the integral receives little contribution at large  $x$. So, to a good approximation, we can take the $D(\omega_i + x/T )$ factor outside the integral and evaluate it at $\omega_i$ and extend the limits of the  integral  to $\pm \infty$ so that
\begin{equation}
    I \approx -T D(\omega_i) \int ^{+\infty}_{-\infty}  \frac{[ \exp{[2ix]}-2\exp{[ix]} +  1]}{x^2} dx
\end{equation}
The reader may be concerned that the large factor $T$ appears on the RHS of the expression defining I in (67), but this concern can easily be allayed by noting that the corresponding sum for the retained terms  is $ \sum_f |\tilde s(\omega_{fi} )|^2 = 2\pi T D(\omega_i) $ and any fractional error in dropping the inter-pulse terms is of order $I/ T$ i.e. nominally of order unity.
The integrand in this approximate expression (67) for $I$ is an analytic function throughout the finite complex plane and the integration contour can be closed in the upper half plane using the infinite semicircle without changing the integral's value. Since the integrand is analytic within the contour the integral's value is zero. So
\begin{equation}
    I \approx 0
\end{equation}
This establishes that the result derived in subsection 4.1, namely
\begin{equation}
\Delta^{(n)}\bigg[\sum_f| c_f|^2 \bigg] =\sum_f|\Delta c_f^{(n)}|^2
\end{equation}
holds for piece-wise constant and contiguous perturbations each of duration $T$. The conditions for validity of this result are
\begin {equation}
 r  \ll \frac{1}{T} \ll \Bigg|\frac{d\omega}{d\ln[D(\omega)]} \Bigg|
 \end{equation}
 where $r$ is the transition rate out of the initial state. This condition has the same form as that for the validity of the conventional derivation of the golden rule, but here, in equation (70), the time is characteristic of the time scale over which the perturbation is sensibly constant (loosely related to the $\gamma$ parameter in (18)) not the range of times for which the result may be applied.

\section{Generalised Golden Rule and Exponential Decay Law}
We have established in the previous section that a train of non-overlapping perturbation pulses of suitably long duration applied to a system previously in a single initial state, cause initial state depletion such that the change in the initial state occupation probability during the $n\text{th}$ pulse is given by 
\begin{equation}
-\Delta^{(n)}\big[|c_i|^2 \big] = \Delta^{(n)}\bigg[\sum_f| c_f|^2 \bigg] =\sum_f|\Delta c_f^{(n)}|^2 = \sum_f\frac{\big|\tilde V_{fi}^{(n)}(\omega_{fi})\big|^2}{\hbar^2}  |c^{(n)}_i |^2
\end{equation}
Introducing  $p_i =  |c_i |^2$, the probability of occupation of the state $i$, and, in particular $p^{(n)}_i =|c^{(n)}_i |^2$, the ( approximately constant ) value of $p_i$ during the $n\text{th}$ pulse, we have the average rate of change of $p_i$ during the  $n\text{th}$ pulse given by 
\begin{equation}
\frac{\Delta^{(n)}p_i}{T} = r^{(n)}  p^{(n)}_i 
\end{equation}
where
\begin{equation}
    r^{(n)} = \frac{1}{T} \sum_f\frac{\big|\tilde V_{fi}^{(n)}(\omega_{fi})\big|^2}{\hbar^2} 
\end{equation}
is the mean transition rate out of the initial state during the $n\text{th}$ pulse and, from (64) is independent of $T$ for large $T$. This result (72) is none other than a finite difference form of the differential equation for exponential decay 
\begin{equation}
    \frac{dp_i(t)}{dt} = - \overline{r}(t) p_i(t)
\end{equation}
with time dependent rate constant $\overline{r}(t)$ and solution 
\begin{equation}
    p_i(t) = p_i(0) \exp{\big[- \int^t_0 \overline{r}(t') dt' \big]}
\end{equation}
where we have used $ \overline{r}(t)$ instead of ${r}(t)$ to distinguish it from the variable transition rate, ${r}(t)$, due to a rising exponential introduced in section 2. Equation (75) is a generalisation of the famous Weisskopf-Wigner ordinary exponential decay result for a constant perturbation and hence constant transition rate.

We emphasize that the adiabatic behaviour displayed in equation (73) has been derived, as mentioned in section 2, without assuming any energy gap in the spectrum of our unperturbed system in contrast to the original derivations of the adiabatic theorem $^{12,13,14}$. 

\section{Weisskopf-Wigner Distilled}
It's instructive at this point to summarize the technique used in the classic paper by Weisskopf and Wigner $^{10}$  on the exponential decay of an excited atom as this is valuable in itself, but also because it is often poorly explained; this is not really surprising when one sees below the detailed care and effort needed to explain the argument in full. One welcome exception to this rule of disappointing explanations  is the excellent treatment of the damped harmonic oscillator in van Kampen's book $^{15}$ . The author strongly recommends the reader to consult this reference especially if they feel the following treatment is in any way unconvincing. 
One starts with the equation of motion (3) without initial approximation save for the assumption that the perturbing potential does not couple final states which is certainly appropriate for their problem. So we start from

\begin{align}
    i \hbar \frac{dc_i(t) }{dt} &= \sum_f V_{if}\exp{[+i \omega_{i,f} t ]}c_f(t) \\
    i \hbar \frac{dc_f(t) }{dt} &=  V_{fi}\exp{[+i \omega_{f,i} t]}c_i(t). 
\end{align}
We assume that the perturbation is switched on at $t=0$ and that the initial conditions are $c_i(0)=1$ and $c_f(0)=0$. We try the solution $c_i(t) = \exp{(-\overline{\omega}t)}$. We do this by first substituting this solution into (77) and finding the implied $c_f(t)$ and then substitute for both $c_i(t)$ and $c_f(t)$ into (76) to check for consistency. Substituting the trial solution $c_i(t) = \exp{(-\overline{\omega}t)}$ into (77) gives the implied $c_f(t)$ as
\begin{equation}
c_f(t) =  \frac{V_{fi}}{\hbar} \frac{1 - \exp{(i\omega_{f,i} -\overline{\omega})t}}{\omega_{f,i}+i\overline{\omega}}.
\end{equation}
We now substitute this implied $c_f(t)$ and the assumed solution
$c_i(t) = \exp{(-\overline{\omega}t)}$ into (76) to check for consistency. We obtain
\begin{equation}
    \overline{\omega} = \frac{i}{\hbar^2} \sum_f |V_{fi}|^2 \frac{\exp{(-i\omega_{f,i} +\overline{\omega})t}-1}{\omega_{f,i}+i\overline{\omega}}.
\end{equation}
Now, generally, this is not consistent because the RHS depends on the time, but it becomes so for small $\overline{\omega}$. To see this we write the consistency condition (79) as 
\begin{equation}
  \overline{\omega} = I_1(t) \exp{(\overline{\omega}t)} +  I_2 
\end{equation}
with
\begin{align}
    I_1(t) &= \int \mathrm{d}\omega_f \frac{f(\omega_f)\exp{(-i\omega_{f,i}t})}{\omega_{f,i}+i\overline{\omega}} \nonumber \\
 \text{and }   I_2 &= -\int \mathrm{d}\omega_f \frac{f(\omega_f)}{\omega_{f,i}+i\overline{\omega}}
\end{align}
and 
\begin{equation}
   f(\omega_f) =  |V_m(\omega_f)|^2 D(\omega_f) .
\end{equation}
In this instance we have allowed the averaged matrix element $V_m$ to depend on the final state energy. It is worth noting at this point that $\overline{\omega}$ may have an imaginary part ( corresponding to a \textit{time independent} energy shift ), and , indeed it does so, as seen in (86) and ( 87) below. But in the following that imaginary part can be considered as incorporated into $\omega_i$ ( with the latter still real) so that the modified $\overline{\omega}$ remains real in (81).  Using, by now, familiar arguments, we find that
\begin{equation}
    I_1(t) = -2 \pi i f(\omega_i) e^{-\overline{\omega}t}
\end{equation}
To evaluate $I_2$ we need to use, valid in the limit of small $\overline{\omega}$,  
\begin{equation}
\frac{1}{\omega+i\overline{\omega}}= P(\frac{1}{\omega}) - i\pi \delta(\omega)
\end{equation}
where $P$ denotes the principal part. So we find 
\begin{equation}
    I_2 =  i \pi f(\omega_i) - P\int \frac{f(\omega_f)}{\omega_{f,i}}\mathrm{d}\omega_f
\end{equation}
so that 
\begin{equation}
 \overline{\omega} = \frac{\pi}{\hbar^2}f(\omega_i) - \frac{i}{\hbar^2}P\int \frac{f(\omega_f)}{\omega_{f,i}}\mathrm{d}\omega_f
\end{equation}
which with (14) and (81) give
\begin{equation}
 \overline{\omega} = r/2 + \frac{i}{\hbar} \Delta E_i
\end{equation}
where 
\begin{equation}
 \Delta E_i = - P \int \frac{|V_m(E_f)|^2 D(E_f) }{E_f - E_i} \mathrm{d}E_f
\end{equation}
is the second order ( and, indeed, the exact ) shift of the energy of the initial state due to the coupling to the final states.
So we have finally
\begin{equation}
 c_i(t)=\exp{(-rt/2)}\exp{(-i\frac{ \Delta E_i t }{\hbar})}
\end{equation}
and
\begin{equation}
 |c_i(t)|^2 =\exp{(-rt)}
\end{equation}
which corresponds to our (75) for the special case of a constant perturbation.

It may be helpful to go through the above steps in more detail as some subtle argument is required to evaluate the integrals $I_1(t)$ and  $I_2$ which we have glossed over. Firstly we have to assume that $f(\omega_f)$ is such that $I_2$ and consequently $I_1(t)$ converge. And we further have to assume that $\omega_i$ is orders of magnitude larger than $\overline{\omega}$; Weisskopf and Wigner point out that for the atomic transitions they are considering, the ratio $\omega_i/\overline{\omega}$ is more than 6 orders of magnitude. We change variable to $x = (\omega_f -\omega_i)/\overline{\omega}$ to obtain 
\begin{equation}
I_1(t) = \int \mathrm{d}x \frac{f(\omega_{i}+\overline{\omega}x )\exp{[-i(\overline{\omega}t)x]}}{x+i} \nonumber \\    
\end{equation}
For sufficiently large values of $\omega_i/\overline{\omega}$, $f(\omega_{i}+\overline{\omega}x )$ will be a slowly varying function of $x$, even for large $x$. At such values of $x$, the factor $(1+x)^{-1}$ will be small and slowly varying and there will be little contribution to the integral when one takes into account the presence of the oscillatory exponential $\exp{[-i(\overline{\omega}t)x]}$. So only the range of  $x$ for which $\overline{\omega}x \ll \omega_{i}$  will contribute significantly to the integral and   $f(\omega_{i}+\overline{\omega}x )$ can be reasonably approximated by $f(\omega_{i})$ and taken outside the integral. Similarly, the limits of integration can be readily extended to $\pm\infty$. So, we can write
\begin{equation}
I_1(t) \approx f(\omega_{i} )\int ^{+\infty}_{-\infty} \mathrm{d}x \frac{\exp{[-i(\overline{\omega}t)x]}}{x+i} \nonumber \\    
\end{equation}
 and,  for $t>0$,  (82) is recovered by contour integration, Jordan's lemma and the residue theorem. One notes that the arguments fail for $t=0$ unless the properties of $f(\omega)$ are particularly favourable. Indeed, even small values of $\overline{\omega}t$ could be problematic in giving non-negligible contribution to the integral $I_1(t)$ at large $x$; here, large $x$ means that the change in $(1+x)^{-1}$ must be small over x changes of $(\overline{\omega}t)^{-1}$.  This helps one understand why we need a different approach for evaluating $I_2$ and cannot just evaluate our result for $-I_1(t)$ at $t=0$.
 
 To evaluate $I_2$ , we need to resolve it into its real and imaginary parts remembering that the imaginary part of $\overline{\omega}$ has been incorporated into $\omega_i$ so that $\overline{\omega}$ can be considered real. For the real part of $I_2$ we have, using $\Delta\omega \gg \overline{\omega}$, and slowly varying $f$, we have,
 
 \begin{align*}
- Re (I_2) &= \int \mathrm{d}\omega_f \frac{\omega_{f,i}f(\omega_f)}{\omega_{f,i}^2+\overline{\omega}^2}\\
&= \int_{|\omega_{fi}| < \Delta\omega} \mathrm{d}\omega_f \frac{\omega_{f,i}f(\omega_f)}{\omega_{f,i}^2+\overline{\omega}^2}+\int_{|\omega_{fi}| > \Delta\omega} \mathrm{d}\omega_f \frac{\omega_{f,i}f(\omega_f)}{\omega_{f,i}^2+\overline{\omega}^2}\\
&\approx f(\omega_i)\int_{|\omega_{fi}| < \Delta\omega} \mathrm{d}\omega_f \frac{\omega_{f,i}}{\omega_{f,i}^2+\overline{\omega}^2}+\int_{|\omega_{fi}| > \Delta\omega} \mathrm{d}\omega_f \frac{\omega_{f,i}f(\omega_f)}{\omega_{f,i}^2+\overline{\omega}^2}\\
&= \int_{|\omega_{fi}| > \Delta\omega} \mathrm{d}\omega_f \frac{\omega_{f,i}f(\omega_f)}{\omega_{f,i}^2+\overline{\omega}^2}\\
&= f(\omega_i) P\int_{|\omega_{fi}| < \Delta\omega} \frac{  \mathrm{d}\omega_f}{\omega_{f,i}}+\int_{|\omega_{fi}| > \Delta\omega} \mathrm{d}\omega_f \frac{\omega_{f,i}f(\omega_f)}{\omega_{f,i}^2+\overline{\omega}^2}\\
&\approx P\int_{|\omega_{fi}| < \Delta\omega} \mathrm{d}\omega_f \frac{ f(\omega_f)}{\omega_{f,i}}+\int_{|\omega_{fi}| > \Delta\omega} \mathrm{d}\omega_f \frac{f(\omega_f)}{\omega_{f,i}}\\
&=  P\int \mathrm{d}\omega_f \frac{f(\omega_f)}{\omega_{f,i}}\\
 \end{align*}
 
 For the imaginary part of $I_2$, we have, again using slowly varying $f$,
 
 \begin{align*}
Im(I_2) &= \int \mathrm{d}\omega_f \frac{\overline{\omega}f(\omega_f)}{\omega_{f,i}^2+\overline{\omega}^2}\\
Im(I_2) &\approx f(\omega_i) \int \mathrm{d}\omega_f \frac{\overline{\omega}}{\omega_{f,i}^2+\overline{\omega}^2}\\
&= \pi f(\omega_i)
\end{align*}
These results for the real and imaginary parts of  $ I_2$  confirm (84).
\\
Hindsight is a wonderful thing. It is possible to simplify the whole argument significantly by noting that $\overline{\omega}$ plays a different role on either side of equation (79). In particular, on the RHS it just plays the role of an infinitesimal and a redundant one at that until one makes the step to equation (80) and evaluates $I_1(t)$ and $I_2$ separately. One notes that there in no singularity in the summand on the RHS of (79) as $\omega_{f} \rightarrow \omega_{i}$ with $\overline{\omega}=0$. So $\overline{\omega}$ can be dispensed with altogether, or the sign reversed. If one takes the latter course, one is still then allowed to separate the integral as in (80). If one reverses the sign of $\overline{\omega}$, then (1) the integral $I_1(t)$ turns out to be zero since the pole has now outside the contour and (2) the imaginary part on the RHS of both (84) and (85) change sign leaving the same result for $\overline{\omega}$ from (80).

\section{Energy Normalisation}
Up to now, at least at the start of each discussion, we have supposed that the spectrum of final states has been , in principle, discrete, albeit quasi continuous, and the initial and final states have been normalised to unity. It is convenient, and even necessary on occasions, to start with a continuous spectrum. We will now show how to do this for the case treated in section 2, an exponentially rising perturbation. We do this for a specific case , scattering by a localised scattering potential in 2 dimensions, but in a homogeneous anisotropic medium. The eigenstates in the absence of the scattering potential are plane waves, but we want to denote these states by   $\psi(E,\phi)$  where $E$ is the energy and $\phi$ the azithmal angle indicating the direction of propagation. The orthonormalisation for these states is chosen to be
\begin{equation}
    \langle \psi(E,\phi) |\psi(E',\phi')\rangle = \delta(\phi - \phi') \delta(E - E')
\end{equation}
It is helpful to relate this normalisation to the more usual one for plane wave states of wave-vector $\boldsymbol{k}$, namely 
\begin{equation}
    \langle \boldsymbol{k} |\boldsymbol{k'} \rangle =  \delta(\boldsymbol{k} -\boldsymbol{k'})
\end{equation}
re-expressing the wave-vector $\boldsymbol{k}$ in terms of its plane polar coordinates magnitude, $k$, and direction, $\phi$, we have

\begin{align}
    \delta(\boldsymbol{k} -\boldsymbol{k'})&=\frac{1}{k}\delta(k -k')\delta(\phi - \phi') \nonumber \\
    &=\frac{1}{k(E,\phi) |\frac{\partial k(E,\phi)}{\partial E}|}\delta(\phi - \phi') \delta(E - E')\nonumber \\
    &=\frac{1}{(2 \pi)^2}\frac{1}{D(E,\phi)}\delta(\phi - \phi') \delta(E - E')
\end{align}
where $D(E,\phi)\delta\phi$ is the density of states per unit area for states with propagation direction within $\delta\phi$ of $\phi$.
From which we see that the density of states is build into the new normalisation : 
\begin{equation}
    |\psi(E,\phi)\rangle = 2\pi\sqrt{D(E,\phi)}|\boldsymbol{k} \rangle 
\end{equation}
   To derive the transition rate induced by the presence of the scattering potential one proceeds in a similar manner to section 2. The state of the system,$\Psi(t)$, is expanded in the eigenstates, $\psi(E,\phi)$, of the unperturbed Hamiltonian
 \begin{equation}
     \Psi(t)=\int \mathrm{d}E \int \mathrm{d}\phi \text{ }c(E,\phi,t) \exp{(-\frac{i}{\hbar}Et) } \psi(E,\phi)
 \end{equation}
  Using initial conditions 
 \begin{equation}
     c(E',\phi',-\infty) = \delta(\phi_i - \phi') \delta(E_i - E')
 \end{equation}
 and with the perturbation approximation, the equation of motion for the final state amplitudes, $c(E,\phi,t)$, is
 \begin{equation}
     i\hbar \frac{dc(E,\phi,t)}{dt} = \langle E,\phi |V| E_i,\phi_i \rangle \\ \exp{[i(E - E_i -i\Gamma)t/\hbar}
 \end{equation}
so that 
 \begin{equation}
    c(E,\phi,t) = - \frac{\langle E,\phi |V|E_i,\phi_i \rangle}{E - E_i -i\Gamma} \exp{[i(E - E_i -i\Gamma)t/\hbar}
 \end{equation}
 leading to
 \begin{equation}
   \frac{d|c(E,\phi,t)|^2}{dt}= \frac{2 \pi}{\hbar} |\langle E,\phi |V|E_i,\phi_i \rangle|^2 \Delta(E - E_i,\Gamma) \exp{(2\Gamma t/\hbar)}
 \end{equation}
 and a transition rate $w(\phi,t)\delta \phi$ into angles within $\delta \phi$  of $\phi$ with 
  \begin{equation}
   w(\phi,t) = \int dE \frac{d|c(E,\phi,t)|^2}{dt}= \frac{2 \pi}{\hbar} |\langle E_i,\phi |V(t)|E_i,\phi_i\rangle|^2 
 \end{equation}
 We see that energy normalisation leads to the density of states factor being incorporated into the matrix element in this simple case. It would appear from this simple example, that energy normalisation does not offer significant advantages: it is just a matter of bookkeeping as to whether the density of states is treated as an extra factor that comes into the calculation as a matter of course or is incorporated into the normalisation at the start. However, there are situations in which energy normalisation has huge advantages. One such is the case of photo-ionisation of an impurity atom in a crystalline medium. In that case the final states are themselves scattering states involving Bloch waves associated from different parts of the bandstructure without apparently any unique density of states. But energy normalisation will incorporate these disparate density of states in the correct ratio with little trouble compared with the general complexity of such photoionisation problems. Another case in which energy normalisation is highly desirable is that in which the final states corresponding to a uniform electric field \textit{in vacuo}. The density of states per unit volume vanishes. One can get correct answers by solving the problem confined to a large box. But it is so much easier with energy normalisation (See Appendix D).
 
 \section{Oppenheimer's Approach : extending \\  the scope of the Golden Rule}
 One way of viewing the time dependent perturbation theory that leads to the Golden Rule is to note that the process of a perturbation, $V$, causing transitions between orthogonal states of the unperturbed Hamiltonian, $H_0$, can also be viewed as the naturally arising transitions between orthogonal, but non-stationary, states of the full Hamiltonian, $H = H_0 + V $ i.e the original Hamiltonian plus perturbation. Oppenheimer $^{15}$ was motivated to take the latter view and extend it to approximately orthogonal non-stationary states, the approximate non-orthogonality arising through spatial separation in his consideration of electric field ionisation of the hydrogen atom. In the cases Oppenheimer considers, the perturbation not only induces transitions, but also creates the continuum of final states!
 
 We start with the full Hamiltonian in the form
 \begin{equation}
     H = T + V_1 + V_2
 \end{equation}
 and assume we can find the orthonormal stationary states and associated energies for either potential on its own :
 \begin{equation}
     H_i = (T+V_i)\psi^{(i)}_{n_i} = E^{(i)}_{n_i} \psi^{(i)}_{n_i} \text{\qquad $i=1,2$}
 \end{equation}
 The essential feature is that the overlap $\langle \psi^{(1)}_{n_2}|\psi^{(2)}_{n_2} \rangle $ is small for eigenstates with energies $E^{(1)}_{n_1} $ and $E^{(2)}_{n_2} $  in the range of interest rather than the potentials $V_1$ and $V_2$ themselves being well separated.
 We start out to derive the appropriate form of the Golden rule for this system by expanding the wavefunction , $\Psi$, of the complete system $ H = T + V_1 + V_2$,  in terms of the eigenfunctions of the two Hamiltonians, $ H_i$ , using the approximation of small overlap,$\langle \psi^{(1)}_{n_2}|\psi^{(2)}_{n_2} \rangle $, to minimise overcompleteness.
 \begin{equation}
     \Psi(t) = \sum_{i=1,2}\sum_{n_i}c^{(i)}_{n_i}\exp{\big[\sfrac{-iE^{(i)}_{n_i}t}{\hbar}}\big]\psi^{(i)}_{n_i}
 \end{equation}
 We substitute this expansion into the time dependent Schr{\"o}dinger equation 
 \begin{equation}
    i\hbar \frac{d\Psi}{dt}= (T + V_1 + V_2)\Psi
 \end{equation}
 to obtain
 \begin{multline}
        i\hbar\sum_{i=1,2}\sum_{n_i}\frac{dc^{(i)}_{n_i}}{dt}\exp{\big[\sfrac{-iE^{(i)}_{n_i}t}{\hbar}}\big]\psi^{(i)}_{n_i} = \sum_{n_1}c^{(1)}_{n_1}\exp{\big[\sfrac{-iE^{(1)}_{n_1}t}{\hbar}}\big]V_2\psi^{(1)}_{n_1} \nonumber \\ +\sum_{n_2}c^{(2)}_{n_2}\exp{\big[\sfrac{-iE^{(2)}_{n_2}t}{\hbar}}\big]V_1\psi^{(2)}_{n_2}
 \end{multline}
 We will suppose that it is the potential $V_2$ that creates the continuum of final states and induces the transitions from an initial eigenstate ,$\psi^{(1)}_{i}$. We will further suppose , as in section 2, that the perturbing potential $V_2(t)$ is switched on in the far past and rises exponentially with initial condition $c^{(1)}_i(-\infty) =1 $ with all other $c^{(i)}_{n_i}$'s zero. But it will always be understood that the eigenstates, $\psi^{(2)}_{n_2}$, used in the expansion of the full wavefunction will be those given by (102), determined with the full strength of $V_2$ i.e.$V_2(0)$. The perturbation approximation then gives one, on taking the scalar product with the basis state  $\psi^{(2)}_{f}$,
 \begin{equation}
     i\hbar\frac{dc^{(2)}_{f}}{dt} \approx \exp{\big[\sfrac{+i(E^{(2)}_{f}-E^{(1)}_{i})t}{\hbar}}\big]\langle \psi^{(2)}_{f}| V_2(t)|\psi^{(1)}_{i} \rangle
 \end{equation}
 If we were to use energy normalisation for the eigenstates of the Hamiltonian, $H_2$, and write for the wavefunction, with $\alpha$ representing any extra parameters beside the energy needed to specify the state, then
\begin{equation}
     \Psi(t) = \sum_{n_1}c^{(1)}_{n_1}\exp{\big[\sfrac{-iE^{(1)}_{n_1}t}{\hbar}}\big]\psi^{(1)}_{n_1} + \int c^{(2)}(E,\alpha)\exp{\big[\sfrac{-iEt}{\hbar}}\big]\psi^{(2)}(E,\alpha)dE
 \end{equation}
 and 
 \begin{equation}
     i\hbar\frac{dc^{(2)}(E,\alpha)}{dt} \approx \exp{\big[\sfrac{+i(E-E^{(1)}_{i})t}{\hbar}}\big]\langle \psi^{(2)}(E,\alpha)| V_2(t)|\psi^{(1)}_{i} \rangle
 \end{equation}
 would replace equations (103) and (104). In analogy with section 2 one obtains the total transition rate as 
 \begin{equation}
   r = \sum_{\alpha}\int dE \frac{d|c(E,\alpha)|^2}{dt}= \frac{2 \pi}{\hbar}\sum_{\alpha} |\langle E_f,\alpha |V_2|E_i\rangle|^2 .
 \end{equation}
 To apply this result to ionisation in an electric field one needs to be able to energy normalise the stationary states in a uniform electric field. It is shown how to do this in Appendix D.\\
  One could, of course , apply the Weisskopf-Wigner method of solution to Oppenheimer's formulation of the ionisation rate for an atom in a uniform electric field. As the reader will recall, the Weisskopf-Wigner solution takes into the energy shift in the initial state energy due to the perturbation responsible for transitions, but this energy shift only relates to the shift due to the interaction with the final states. No account is taken of the change in the initial state energy level due to the the other initial system (1) states and $V_2$. In tunnelling problems, such as ionisation in an electric field, such a shift in energy could have a marked influence on the tunneling rate. One way to improve on the Weisskopf-Wigner solution, or, indeed, any other solution, is as follows. If, in the first instance, one uses $V_1$ as the potential responsible for the initial bound state and $V_2$ as the applied electric field one could allocate the part of $V_2$ that overlaps with $V_1$ to $V_1$ itself. Then any Stark shift will be incorporated into the initial state. So we can incorporate this effect in our result (107) above by just incorporating the Stark effects in $E^{(1)}_{i}$ and $\psi^{(1)}_{i} $.
  
  \section{ Summary}
  The reworking of a derivation of the Golden Rule of time dependent perturbation theory as used in formal scattering theory has proved remarkably fertile ground for nurturing one's understanding thereof. In the formal theory of scattering one considers the adiabatic turning on of the scattering potential : the static potential is multiplied by a rising exponential $e^{\gamma t }$ and at the end of the calculation of the transition rate the limit $\gamma \rightarrow 0$ is taken. Taking this limit literally is of little practical use as the perturbation must tend to zero as well if the initial state amplitude is to remain close to unity. But if one does not take this limit , but instead just considers $\gamma$ sufficiently small, then one finds one can derive the standard result for the transition rate which is valid for all negative times. There is no need to put a `Goldilocks' type of restriction on the time interval for which the result is valid: the time interval must not be too short or the constant transition rate will not be established; the time interval must not be too long otherwise the perturbation theory will not longer be valid. While one finds a `Goldilocks' type of restriction on our derivation's validity is needed, these restrictions are all in the frequency/energy domain. Another remarkable feature is that when one achieves approximate energy conservation one also achieves a quasi adiabatic following by the transition rate of the square of the perturbing potential, which varies as $e^{2\gamma t }$, with a prefactor independent of $\gamma$ ; and similarly for the wavefunction as one would expect. The system seems to have lost all memory of the behaviour of the perturbation at previous times. This immediately raises the question as to whether this result is more general or is just restricted to a rising exponential.
  
  We have explored this question of generalisation first by looking at variations on the rising exponential : linear combinations thereof; a time varying perturbation in the form of a pulse with rising and falling exponentials for the leading and trailing edges respectively. The results all suggest the generalisation is true. We establish the generalisation by considering the time dependence of the perturbation as a succession or train of  non-overlapping pulses with their centres separated by a suitably long time interval. We find that the overall effect on the depletion of the initial state of these perturbative pulses is simply additive with the only influence of previous pulses is on the occupation probability of the initial state. In particular we find that these pulses can be rectangular with no gap in between, so that the perturbing potential can be piecewise constant. This then allows us to derive the Golden rule suitable for slowly varying perturbing potentials and in particular derive a generalisation of the Weisskopf-Wigner exponential decay result i.e. exponential decay of an excited state with a time varying decay constant.
  We finish by distilling the essence of the classic, but difficult, papers by Weisskopf-Wigner on the decay of an excited atom by radiative emission and Oppenheimer on the electric field ionisation of an atom.
  
  The results obtain here, along with those obtained some years ago by the author $^{2,3}$ on adiabatic following, albeit obtained only in perturbation theory, strongly suggest that the energy gap condition used for proving the general adiabatic theorem in quantum mechanics $^{12,13,14}$ is not necessary and indeed a proof of the theorem for a continuous spectrum has appeared $^{17}$.
  
  As a parting comment, it is as well to summarise the difference between the conventional derivation of the Golden Rule of time dependent perturbation theory and that presented here in section 2. In the conventional derivation the constant perturbation is switched on abruptly and, after a while, approximate energy conservation is achieved along with a constant transition rate. In the derivation presented here one is not restricted to a constant perturbation. The variation in the perturbing potential needs only to be slow enough for approximate energy conservation to be maintained, a mild restriction in practice. The system responds quasi adiabatically, the wavefunction being an approximate eigenstate at all times of the instantaneous total Hamiltonian. This entails a mixture of final states of the unperturbed Hamiltonian which leads to a transition rate dependent only on the instantaneous value of the  perturbation. The approximate conservation of energy is expressed in terms of a physically appealing Lorentzian line shape. Indeed, the derivation presented here provides not only one of wider applicability, but also one of greater physical insight.

\section*{ References}

\begin{enumerate}
\item L. I. Schiff Quantum Mechanics ( 2nd edition ) McGraw-Hill (1955) Ch VIII 
\item M.G.Burt Semicond. Sci. Technol. 8 (1993) 1393
\item M.G.Burt in 'Coherent Optical Interactions in Semiconductors'\\ (R.T.Phillips Ed. ) Plenum (1994) p 273  
\item J. J. Sakurai Modern Quantum Mechanics (1994) Addison-Wesley p332 
\item E Fermi Nuclear Physics University of Chicago Press 1950 pp 75, 136, 142 and 148
\item P. A. M. Dirac Proc. Roy. Soc. A114 243 (1927) 
\item P. A. M. Dirac The Principles of Quantum Mechanics (4th edition ) OUP 1958 pp 178 - 181
\item T. Visser Am. J. Phys 77, 487 (2009)
\item E. Merzbacher Quantum Mechanics Wiley ( 1961) pp 482 – 486 and  ( 3rd edition) Wiley (1999) pp 517 - 520
\item V.F. Weisskopf and E.P. Wigner Z. Physik 63 (1930) 54 
\item J.R. Oppenheimer Phys Rev 31 66 (1928) 
\item M. Born and V. Fock Zeit. f. Phys. 51 ( 1928) p165
\item T. Kato Journ. Phys. Soc. Japan 5 (1950) 435
\item A Messiah 1999 Quantum Mechanics (Dover) Chapter XVII pp 744 - 750
\item N. G. van Kampen Stochastic Processes in Physics and Chemistry Elsevier( 3rd edition ) (2007) pp 428-436 
\item J R Oppenheimer Phys Rev 31  ( 1927 ) 66
\item J. E. Avron and A. Elgart (1999). "Adiabatic Theorem without a Gap Condition". Communications in Mathematical Physics. 203  445–463. or arXiv:math-ph/9805022
\item see e.g. D.E.Aspnes Phys Rev 147, 554 (1966)
\item see e.g. M Cardona Modulation Spectroscopy pp 325 – 327 Academic, New York
and London (1969)

\end{enumerate}
\section *{Acknowledgements}

I am indebted to Dr D T Cornwell for interesting and stimulating conversations that reawakened my interest in this topic. I am also grateful to Dr D.T.Cornwell, Prof R.A.Abram and Dr B.A.Foreman for detailed, perceptive and helpful comments on various parts of earlier versions of the manuscript. I would also like to thank Prof S.J.Clark for his interest, support and constructive comments.

\section*{Appendix A : Averaged  Squared Modulus of the Matrix Element }

In the derivation of the Golden Rule in section 2 we simplified matters, in a nonessential way , by taking the matrix element of the perturbation, $V_{if}$ , between the initial state, $i$, and any final state, $f$, as independent of  both $i$ and $f$. It will be helpful to generalize this, because there are situations in which this is not so, and it is useful to see how the overall form of the Golden rule, as stated in equations (13) and (14), is retained. For the transition rate to the state ,$f$, given in equation (7) we have the dependence
\begin{equation}
    |V_{fi}|^2 \Delta ( E_{fi}, \Gamma )
\end{equation}
on $f$. We imagine , to keep things simple, that the final state , besides its energy, $E$ , needs just a single discrete parameter , $\sigma$ , to define it. Leaving the dependence of $V_{if}$ on $i$ implicit , we write
\begin{equation}
    V_{fi} = V_m(E,\sigma)
\end{equation}
Even in one dimensional scattering problems the presence of the parameter, $\sigma$ , is necessary since, in general, the matrix element will depend on whether the final state corresponds to forward or backward scattering and not just on the energy alone. We now suppose that $\Gamma$ is sufficiently small that $V_m(E,\sigma)$ is approximately constant over the line shape , $\Delta$ . As before we also assume that $\Gamma$ is sufficiently small that the final state energy levels, for each value of $\sigma$ , are uniformly distributed over the line shape  with density, $D(E, \sigma )$ , so that

\begin{align}
    \sum_{f}|V_{fi}|^2 \Delta ( E_{fi}, \Gamma )
    &= \sum_{\sigma}|V_m(E_i ,\sigma)|^2 D(E_i, \sigma ) \nonumber \\
    &= \frac{\sum_{\sigma}|V_m(E_i ,\sigma)|^2 D(E_i, \sigma )}{\sum_{\sigma}D(E_i, \sigma )}  \sum_{\sigma}D(E_i, \sigma )\nonumber\\
    &= \overline {|V_m(E_i)|^2} D(E_i) \nonumber \\
\end{align}
with $D(E_i)$ , the total density of states, being given by 
\begin{equation}
    D(E_i) = \sum_{\sigma}D(E_i, \sigma )
\end{equation}
and the average,$\overline {|V_m(E_i)|^2} $, of the squared modulus of the matrix element given by 
\begin{equation}
    \overline {|V_m(E_i)|^2}  = \frac{\sum_{\sigma}|V_m(E_i ,\sigma)|^2 D(E_i, \sigma )}{\sum_{\sigma}D(E_i, \sigma )} 
\end{equation}
It might be commented that having the density of states dependent on $\sigma $ for a one dimensional problem is superfluous, since for free particles the density of states is the same for both forward and backward scattering, but there examples in solid state physics in which, even though only one extra parameter, $\sigma $, is needed, besides the energy, to specify the final state, the density of states will still depend on such a parameter.

Perhaps a more concrete example is the scattering of an electron in a two dimensional crystal by a localised defect potential. If all the final states are within a single band , then the final state can be specified by its energy $E$ and the direction of the wavevector which we can specify with an angle, $\phi$ . The matrix element may be written
\begin{equation}
    V_{fi} = V_m(E,\phi)
\end{equation}
The density of states, $ D(E,\phi)$ , is such that the number of states within $\delta E$ of $E$ and within $\delta \phi$ of $\phi$ is $ D(E,\phi) \delta E \delta \phi$ . The sum over states is evaluated in an analagous way to the previous one dimensional case:
\begin{align}
    \sum|V_{fi}|^2 \Delta ( E_{fi}, \Gamma )
    &= \int|V_m(E_i ,\phi)|^2 D(E_i, \phi ) \mathrm{d} \phi \nonumber \\
    &= \frac{\int|V_m(E_i ,\phi)|^2 D(E_i, \phi )\mathrm{d} \phi}{\int D(E_i, \phi ) \mathrm{d} \phi}  \int D(E_i, \phi ) \mathrm{d} \phi\nonumber\\
    &= \overline {|V_m(E_i)|^2} D(E_i) \nonumber \\
\end{align}
with $D(E_i)$ , the total density of states, being given by 
\begin{equation}
    D(E_i) = \int D(E_i, \phi ) \mathrm{d} \phi
\end{equation}
and the average,$\overline {|V_m(E_i)|^2} $, of the squared modulus of the matrix element given by 
\begin{equation}
    \overline {|V_m(E_i)|^2}  = \frac{\int |V_m(E_i ,\phi)|^2 D(E_i, \phi )\mathrm{d} \phi}{\int D(E_i, \phi )\mathrm{d} \phi} 
\end{equation}

\section*{Appendix B : Derivation of the Golden Rule with a harmonic perturbation}

In section 2 we derived the golden rule for a constant perturbation using a rising exponential envelope as a device to make the derivation perspicuous. We show here that it is a relatively straightforward matter to extend the derivation to a harmonic perturbation. We take the perturbation to be 

\begin{equation}
    V(t)=V \exp(\gamma t ) \big[ ( \exp( -i \omega t ) +  \exp( +i \omega t ) \big ]
\end{equation}
with $\omega \gg \gamma $.

Integrating the equation for $c_f(t)$ in first order, as in section 2, we find
\begin{equation}
 c_f(t) \approx \frac{-i}{\hbar}  V_{fi}  \exp(\gamma t )  \bigg[ \frac{\exp{(+i( \omega_{f,i} - \omega ) t) }}{+i (\omega_{f,i} - \omega)+\gamma } + \frac{\exp{(+i( \omega_{f,i} + \omega )t  ) }}{+i (\omega_{f,i} + \omega)+\gamma } \bigg ]
\end{equation}
and forming the modulus squared we find

\begin{equation}
 |c_f(t)|^2 \approx \frac{|V_{fi}|^2}{\hbar^2} \exp(2 \gamma t )  \bigg[ \frac{1}{ (\omega_{f,i} - \omega)^2+\gamma^2 } + \frac{1}{ (\omega_{f,i} + \omega    )^2+\gamma^2 } \bigg ]
\end{equation}
where we have neglected two terms within the square brackets in equation (20), namely
\begin{equation}
   \frac{\exp{(\pm 2i  \omega  t) }}{(\gamma \mp i (\omega_{f,i} - \omega  ))(\gamma \pm i (\omega_{f,i}  +  \omega )) }  
\end{equation}

There are two factors which taken together make this an extremely good approximation. Firstly , if we take the ratio the values of the moduli of the discarded and retained terms at their respective maxima, we find this is of order $\gamma / 2 \omega$ . Secondly, the exponentials $ \exp{(\pm 2i  \omega  t)} $ vary rapidly relative to the envelope $\exp(2 \gamma t )$ so that the terms have an average value close to zero. We now find from equation (20) that the transition rate to a single final state $f$ is 
\begin{equation}
 \frac{d |c_f(t)|^2 }{dt} \approx \frac{2 \pi}{\hbar}  |V_{fi}|^2 e^{2\gamma t} \bigg[ \Delta ( E_{fi}-E, \Gamma ) + \Delta ( E_{fi}+E, \Gamma ) \bigg]
\end{equation}

The derivation of the total transition rate mirrors that for the constant perturbation case. 

\section*{ Appendix C : Further Evidence of Adiabatic  Following }

We saw in the previous section that for a perturbation turned on exponentially, and sufficiently slowly, the transition rate follows the the squared modulus of the matrix element. We now demonstrate further evidence that this result is of wider validity than the simple case of a single exponential time dependence : we consider the case in which the time dependence of the perturbation is a superposition of exponentials; we will see that adiabatic following is still displayed.

We suppose that the perturbation is given by
\begin{equation}
V(t) = \int \overline{V}(\gamma) \exp{(\gamma t)} \mathrm{d} \gamma
\end{equation} 
where $\overline{V}(\gamma) $ is only non zero for those values of $\gamma$ that obey the condition $(10)$ required to ensure adiabatic following for the single exponential case.The result corresponding to $(6)$, for the final state amplitude, $c_f(t)$ , is  
\begin{equation}
 c_f(t) \approx - \exp{(+i \omega_{f,i} t ) }  \int \mathrm{d} \gamma \overline{V}_m (\gamma)\frac{\exp{( \gamma t) }}{ E_{fi} - i \Gamma } 
\end{equation}
Again, to reduce unnecessary complications, we assume that $\overline{V}_{fi}$ is independent of both $i$ and $f$ and write it as $\overline{V}_m$. The time rate of change of the probability of finding the system in state $f$ is given by
\begin{multline}
  \frac{d |c_f(t)|^2 }{dt} \approx \\  
   \iint \mathrm{d} \gamma_1 \mathrm{d} \gamma_2 \overline{V}_m (\gamma_1) \overline{V}_m (\gamma_2)^* \frac{\gamma_1 + \gamma_2}{ (E_{fi} - i \Gamma_1)(E_{fi} + i \Gamma_2) } \exp{[(\gamma_1 + \gamma_2) t] }
\end{multline}
The total transition rate out of the initial state is then
\begin{multline}
\iint \mathrm{d} \gamma_1 \mathrm{d} \gamma_2 \overline{V}_m (\gamma_1) \overline{V}_m (\gamma_2)^* \exp{[(\gamma_1 + \gamma_2) t]} \times \\ \int\mathrm{d}E_fD(E_f) \frac{\gamma_1 + \gamma_2}{ (E_{fi}-i\Gamma_1)(E_{fi}+i \Gamma_2) } 
\end{multline}

The integrand of the integral with respect $E_f$ will be sharply peaked about $ E_f \approx E_i$ if the $\Gamma$'s in (126) are sufficiently small to obey the condition (10). Then $D(E_f)$ can be evaluated at $E_f = E_i $ and taken outside the integral w.r.t $E_f $ and that integral becomes
\begin{equation}
\frac{D(E_i)}{\hbar} \int\mathrm{d}E_f\frac{\Gamma_1 + \Gamma_2}{ (E_{fi} - i \Gamma_1)(E_{fi} + i \Gamma_2) } 
\end{equation}
The integral in (127) can be evaluated by contour integration by closing in either the lower or upper half plane to give $2\pi$ and the transition rate, $w$, is found to be
\begin{equation}
\frac{2\pi}{\hbar} \Bigg[\iint \mathrm{d} \gamma_1 \mathrm{d} \gamma_2 \overline{V}_m (\gamma_1) \overline{V}_m (\gamma_2)^* \exp{[ (\gamma_1 + \gamma_2) t] } \Bigg ] D(E_i)
\end{equation}
or
\begin{equation}
w(t) \approx \frac{2\pi}{\hbar} |\overline{V}_m (t)|^2  D(E_i)
\end{equation}
which again displays quasi adiabatic following and suggests that the result is more general.

\section*{Appendix D : Energy Normalisation for the Wavefunction of a Charged Particle in a Uniform Electric Field }
 We start with the Schr{\"o}dinger equation in one dimension for a particle in a uniform field, $F$, in the positive $x$ direction.
 \begin{equation}
     -\frac{\hbar^2}{2m} \frac{d^2\Psi}{dx^2}-Fx\Psi = E \Psi
 \end{equation}
 With change of variable to 
 \begin{equation}
     \xi = -\frac{1}{a}(x + \sfrac{E}{F}) \text{\qquad in which \qquad } a = \bigg[\frac{\hbar^2}{2mF} \bigg]^{1/3}
 \end{equation}
 we obtain Airy's equation 
 \begin{equation}
    \frac{d^2\Psi}{d\xi^2}-\xi\Psi =0
 \end{equation}
For $\Psi$ finite everywhere we have 
\begin{equation}
    \Psi  \propto Ai( \xi) = Ai(-\frac{1}{a}(x + \sfrac{E}{F}))
\end{equation}
where $Ai( \xi)$ is Airy's function which has integral representation$^{17,18}$
\begin{equation}
 Ai(\xi) = \int ^{+\infty}_{-\infty}\frac{ds}{2\pi}\exp{[is^3 + i\xi s]}
\end{equation}
The normalisation integral can be evaluated using this integral representation$^{18}$:

\begin{multline}
  \int ^{+\infty}_{-\infty} d\xi Ai(\xi-\xi_1)Ai(\xi-\xi_2) = \\
  \int ^{+\infty}_{-\infty} d\xi  \int^{+\infty}_{-\infty} \frac{du}{2\pi}\exp{[iu^3 + i(\xi-\xi_1)u]} \int ^{+\infty}_{-\infty}\frac{dv}{2\pi}\exp{[iv^3 + i(\xi-\xi_2) v]} \\ =
  \int^{+\infty}_{-\infty} \frac{du}{2\pi}\exp{[iu^3 -i\xi_1)u]} \int ^{+\infty}_{-\infty}\frac{dv}{2\pi}\exp{[iv^3 - i\xi_2 v]} \text{\quad} 2\pi\delta(u+v) \\ = \delta(\xi_1-\xi_2)
\end{multline}
So if we write
\begin{equation}
    \Psi ( x , E )  = \frac{1}{a\sqrt{F}}Ai[-\frac{1}{a}(x + \sfrac{E}{F})]
\end{equation}
we find $\Psi$ is energy normalised i.e.
\begin{equation}
   \int ^{+\infty}_{-\infty}dx \Psi ( x , E )^* \Psi ( x , E' ) = \delta(E - E')
\end{equation}
We could check this result, and the author has indeed checked it, by putting an impenetrable barrier at a large distance down field to produce a discrete, albeit quasi continuous, spectrum and computing the density of states. It is a tedious long-winded business, even for this simple case, but certainly demonstrates the power of energy normalisation and that not all density of states are proportional to the size of the system at large sizes

\end {document}